\documentclass[longbibliography,twocolumn,amsmath,amssymb,amsfonts,superscriptaddress,floatfix,showpacs]{revtex4-1}
\usepackage[utf8]{inputenc}
\usepackage[english]{babel}
\usepackage[colorlinks=true,citecolor=green,linkcolor=red,urlcolor=blue]{hyperref}
\usepackage{braket}
\usepackage{bbm}
\usepackage{graphicx}
\usepackage{color}
\usepackage{dsfont}
\usepackage{float}

\newcommand{\skippart}[1]{}
\newcommand{\simplesection}[1]{\emph{#1} ---}

\begin{document}

\title{Galois conjugated tensor fusion categories and non-unitary CFT}
\author{Laurens~\surname{Lootens}}
\affiliation{Department of Physics and Astronomy, Ghent University, Krijgslaan 281, S9, B-9000 Ghent, Belgium}
\author{Robijn~\surname{Vanhove}}
\affiliation{Department of Physics and Astronomy, Ghent University, Krijgslaan 281, S9, B-9000 Ghent, Belgium}
\author{Jutho~\surname{Haegeman}}
\affiliation{Department of Physics and Astronomy, Ghent University, Krijgslaan 281, S9, B-9000 Ghent, Belgium}
\author{Frank~\surname{Verstraete}}
\affiliation{Department of Physics and Astronomy, Ghent University, Krijgslaan 281, S9, B-9000 Ghent, Belgium}
\affiliation{Vienna Center for Quantum Technology, University of Vienna, Boltzmanngasse 5, 1090 Vienna, Austria}

\begin{abstract}

We provide a generalisation of the matrix product operator (MPO) formalism for string-net projected entangled pair states (PEPS) to include non-unitary solutions of the pentagon equation. These states provide the explicit lattice realisation of the Galois conjugated counterparts of (2+1) dimensional TQFTs, based on tensor fusion categories. Although the parent Hamiltonians of these renormalisation group fixed point states are non-Hermitian, many of the topological properties of the states still hold, as a result of the pentagon equation. We show by example that the the topological sectors of the Yang-Lee theory (the non-unitary counterpart of the Fibonacci fusion category) can be constructed even in the absence of closure under Hermitian conjugation of the basis elements of the Ocneanu tube algebra. We argue that this can be generalised to the non-unitary solutions of all $SU(2)$ level $k$ models. The topological sector construction is demonstrated
 by applying the concept of strange correlators to the Yang-Lee model, giving rise to a non-unitary version of the classical hard hexagon model in the Yang-Lee universality class and obtaining all generalised twisted boundary conditions on a finite cylinder of the Yang-Lee edge singularity. 
\end{abstract}
\maketitle

\simplesection{Introduction}
Tensor fusion categories~\cite{etingof2015tensor}  describe non-local ``symmetries'' in both topological quantum field theories (TQFTs)~\cite{Witten1988,atiyah1988topological} and conformal field theories (CFTs)~\cite{Moore1989,segal1988definition}. On the one hand, lattice systems with non-chiral topological quantum order can be realised by the Turaev-Viro state sum constructions~\cite{Turaev1992,barrett1996invariants} or by string-net wavefunctions~\cite{Levin2005}. In these descriptions the topological order emerges from the condensations of ``strings'' (Wilson lines) that do not necessarily fuse according to group rules but more generally according to an algebra. The topological sectors (emerging anyonic excitations), given by the Drinfeld center of the input fusion category~\cite{drinfeld,JOYAL199143,majid1991representations,DrinfeldCenter}, are practically found by block-diagonalising the Ocneanu tube algebra~\cite{ocneanu1994chirality,tubealgebra,evans1998quantum}. On the other hand, classical two dimensional critical systems exhibiting a second order phase transition are captured in the scaling limit by (1+1) dimensional (rational) CFTs. This connection between TQFTs and CFTs was first envisioned by Witten and developed by Fr\"ohlich, Fuchs, Runkel and Schweigert~\cite{Felder2000,Fuchs2002,Frohlih2004} by means of tensor fusion categories. The relation between topological conformal defects in CFTs and topological sectors in TQFTs was shown in the latter works and explicitly established on the lattice using string-net models~\cite{Aasen2016}.

In former work~\cite{vanhove2018mapping} this connection was further developed using the concept of strange correlators (SCs), originally used as a detection mechanism for short-range entangled symmetry protected topological (SPT) phases~\cite{You2014}. The SC establishes a map from a (2+1)- dimensional quantum state $\ket{\Psi}$ to a corresponding classical two-dimensional partition function by means of taking the overlap of the state with an unentangled state $\ket{\Omega}$. We argued that the SC construction for long-range entangled string-net wavefunctions in terms of PEPS was useful because it systematically describes the non-local symmetries of the topological properties of emergent CFTs~\cite{Aasen2016} in terms if matrix product operators (MPOs). Moreover the SC allows for the construction of the conformal blocks out of the topological sectors of the string-net wavefunction, which is typically hard to do \cite{Mong2014,Aasen2016,Hauru2016}. The conformal spins of the emergent CFT are in one-to-one correspondence to the topological spins of the original TQFT~\cite{Moore1989,kitaev2006anyons}.

The generalised SC construction is a Euclidean counterpart to the anyonic chain models~\cite{PhysRevLett.98.160409,PhysRevLett.103.070401,Buican2017}, in which topological defects have already been considered in the context of generalised twisted CFT partition functions for both the Fibonacci chain~\cite{PhysRevLett.98.160409} and the non-unitary Yang-Lee Galois conjugated version~\cite{ardonne2011microscopic}. However, the identification of the topological sectors in the spectra in the non-unitary case has not been discussed, to our knowledge, in the case of a non-trivial twist in the partition function on a torus~\cite{Petkova2000,Petkova2001}. This requires the full Ocneanu tube algebra as basis elements for the projection onto all topological superselection sectors. We stress that the (1+1)-dimensional quantum Hamiltonian has the same MPO symmetry as the SC transfer matrix, since the quantum Hamiltonian can also be constructed out of a physical projection of string-net PEPS tensors, hereby promoting the virtual degrees of freedom in the (2+1) dimensional model to physical degrees of freedom in the quantum Hamiltonian and preserving the MPO symmetry.

We aim in this work to demonstrate that the MPO-formalism that was developed in~\cite{Sahinoglu2014,williamson2014matrix,Bultinck2017,Williamson2017} can be extended to non-unitary solutions of the pentagon equation given some constraints on the tensor fusion categories that are considered. At the heart of this demonstration lies the pentagon equation and more particularly the existence of non-unitary and self-inverse F-symbols that are solutions of it. The Projected entangled pair states (PEPS) that are built out of these F-symbols form a natural representation of string-net wavefunctions with non-local symmetries described by MPOs. These MPOs naturally capture the topological order of the phase. However, because of non-unitarity, the Hermitian conjugate of these MPOs bares no relationship to the MPOs associated with the dual objects of the category and form a seperate algebra instead. The topological sectors that are given by the idempotents of the Ocneanu tube algebra~\cite{ocneanu1994chirality,tubealgebra,evans1998quantum} are identified with the emergent anyonic excitations in the theory. Since the tube algebra is no longer closed under Hermitian conjugation, it does not form a $C^{*}$-algebra and therefore the existence of the idempotents is not guaranteed. It is however possible to construct such idempotents for braided mudular tensor categories (BMTCs), as these theories have invertible S-matrices and a braid matrix $R$. This is a consequence of a simple generalization of the explicit construction of the idempotents based on the S matrix and the solution of the hexagon equation carried out in \cite{koenig2010quantum} to non-unitary solutions. For the formalism to work, we further need theories where particles are identified with their anti-particles. Translated to the fusion rules of the BMTC, this means that the identity element of the algebra is present in the fusion product of every element with itself. The $\text{su}(2)_k$ fusion rules obey this property and the associated tensor fusion category is braided and modular. We apply the SC construction to the example of $\text{su}(2)_3$ by exactly diagonalising the obtained transfer matrix of the non-unitary version of the hard-hexagon model. This model falls into the Yang-Lee universality class with central charge $c = -22/5$ and the non-trivial Fibonacci MPO corresponds to the topological conformal defect with conformal weight $h = -1/5$ \cite{DiFrancesco1997}. Finally, the conformally twisted spectra are projected onto the topological sectors, by means of the obtained expressions for the idempotents, to illustrate their correspondence with the sesquilinear combination of the conformal characters on a torus~\cite{vanhove2018mapping,Petkova2000,Petkova2001}.

\simplesection{String nets and the pentagon equation}
The tensor network representation of string-net models on a hexagonal lattice can be constructed by defining the following PEPS tensors out of which the hexagonal lattice can be built \cite{buerschaper2009explicit}:
\begin{equation}
\label{eq1}
\includegraphics[width=0.9\linewidth]{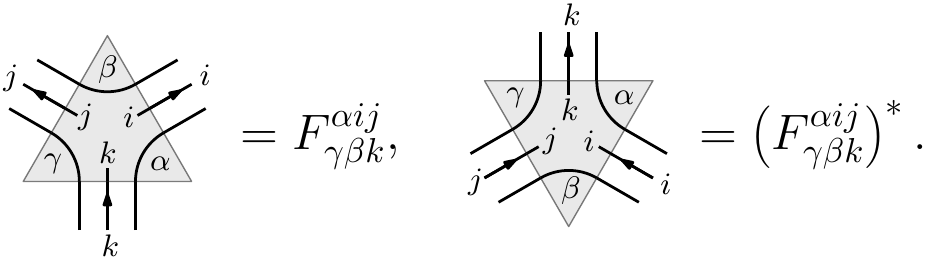}
\end{equation}
Here, $F^{abc}_{def}$ denotes the $F$ symbol of the input fusion category, also known as the quantum $6j$ symbol. The non-local MPO symmetries of these string-net models are required to satisfy the local pulling-through equation \cite{Sahinoglu2014,williamson2014matrix,Bultinck2017,Williamson2017}:
\begin{equation}
\label{eq2}
\includegraphics[width=0.7\linewidth]{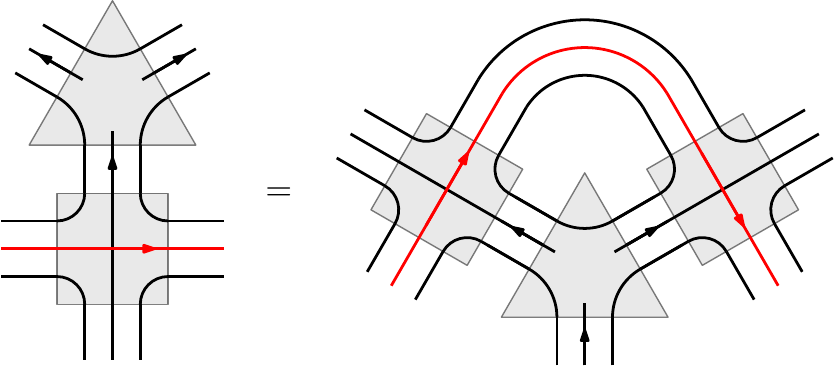}
\end{equation}
and a similar equation for the second PEPS tensor in eq. \ref{eq1}. The relative directions of the black and red arrows denotes the handedness of the MPO; by convention we call the above MPOs left-handed. If we now define these MPOs as
\begin{equation}
\label{eq3}
\includegraphics[width=0.5\linewidth]{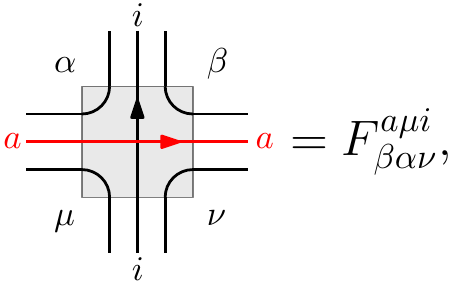}
\end{equation}
we find that the pulling-through property in eq. \ref{eq2} is nothing more than the pentagon equation, which is known to envelop all possible consistency equations and therefore completely determine the $F$-symbols, a property known as Maclane coherence theorem in category theory \cite{kitaev2006anyons}. Ocneanu rigidity then states that any small deformation of a solution to the pentagon equation can be absorbed through a gauge transformation on the $F$-symbols, so that the solutions fall into discrete families. The MPOs we defined above carry the label $a$, and they form a closed algebra that corresponds to the input fusion category of the string-net model. From a tensor network point of view, the fundamental theorem for MPS (which can be trivially extended to MPOs) implies the existence of an object that implements the fusion of two MPOs:
\begin{equation}
\includegraphics[width=0.9\linewidth]{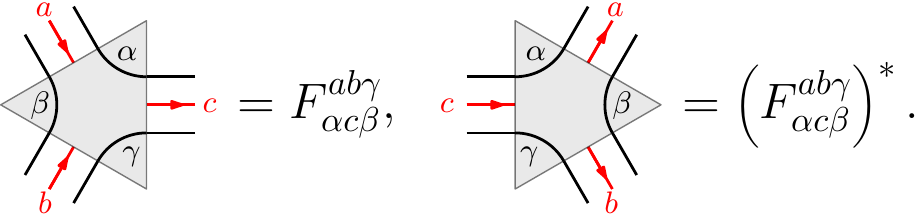}
\end{equation} 
The definitions of the above tensors assume these $F$ symbols to be unitary: $\sum_f F^{abc}_{def} \left(F^{abc}_{de'f}\right)^* = \delta_{ee'}$, such that in particular the fusion of two MPOs is an isometry. In this work, we want to lift this unitarity restriction, for it is known that the pentagon equation also admits non-unitary solutions. Solving the pentagon equation is highly non-trivial, but there are algorithms available that are able to find all solutions, be it unitary or not. For certain simple types of fusion categories however, based on representations of quantum groups, there exist analytic expressions for the $F$-symbols \cite{ardonne2010clebsch}, and it is these models that we will consider in the following sections.

\simplesection{Galois conjugates of $\text{su}(2)_k$ models}
The fusion categories we consider in this work are quantum deformations of $SU(2)$, where we limit the number of representations to $k+1$ (generalised) angular momenta that take the values $j = 0,\frac{1}{2},1,\dots,\frac{k}{2}$. These fusion categories are denoted as $\text{su}(2)_k$. In the language of anyons, taking the tensor product of two representations of $\text{su}(2)_k$ corresponds to the fusion of two anyons, where the fusion rules are dictated by the usual $SU(2)$ tensor product structure:
\begin{equation}
j_1 \times j_2 = \sum_{j_3 =|j_1 - j_2|}^{\min{(j_1 + j_2,k - j_1 - j_2)}} j_3
\end{equation}
with the modification that $j_3$ is also bounded by $k - j_1 - j_2$ to ensure associativity of the fusion rules. The explicit solutions to the pentagon equation for all $\text{su}(2)_k$ fusion categories have been found using quantum group techniques \cite{kirillov1988representations}. It turns out that the $F$ symbols for all these theories depend on a single deformation parameter $q$, which is a root of unity given by $q = e^ {p\cdot 2\pi i/(k+2)}$ where the integer $p$ runs from $1$ to $(k + 2)/2$ and $p$ and $k +2$ are coprime. From this we see that for all levels $k$, $p = 1$ gives one possible value for $q$, and it is this choice of $q$ that gives unitary $F$ symbols. Starting at $k = 3$, there is more than one allowed value of $p$ and these $F$ symbols are non-unitary. The process of going from one root of unity to another is referred to as Galois conjugation, and it is these Galois conjugated non-unitary theories that are the subject of this paper. One important property of the unitary $F$-symbols in these $\text{su}(2)_k$ models is that they are real, making complex conjugation unnecessary in the defenitions of the PEPS and fusion tensors. The unitarity condition for unitary $\text{su}(2)_k$ models simply becomes the statement that the $F$-symbols, when interpreted as matrices $\left[F^{abc}_{d}\right]^e_f$, are involutory matrices. This statement continues to hold for the Galois conjugated non-unitary categories, and we can exploit this property to soften the unitarity condition: we propose to define the PEPS, MPO and fusion tensors for the non-unitary categories without introducing complex conjugation. Galois conjugation preserves the algebraic properties of the F-symbols, so we expect all the properties of the unitary categories to carry over to the non-unitary ones.

The simplest example of a non-unitary fusion category occurs at level $k=3$, where we have $4$ representations $\{0,\frac{1}{2},1,\frac{3}{2}\}$. The fusion rules can be more conveniently expressed as $\text{Fib} \otimes \mathds{Z}_2$, where $\text{Fib}$ denotes the category with two objects $\{\mathbf{1},\mathbf{\tau\}}$ with  $\mathbf{\tau}\times\mathbf{\tau} = \mathbf{1} + \mathbf{\tau}$ as the only non-trivial fusion rule. The $F$-symbols of a fusion category where the fusion rules factorise into a tensor product are given by the product of the $F$ symbols of the factors in the tensor product. We will restrict ourselves to the subalgebra of integer representations of $\text{su}(2)_3$ with trivial $\mathds{Z}_2$ charge, which have trivial $\mathds{Z}_2$ $F$-symbols. The only non-trivial $\text{Fib}$ $F$-symbols are then given, in terms of the $q$ deformation parameter, by
\begin{equation}
F^{\tau\tau\tau}_{\tau e f} = 
\begin{pmatrix}
\frac{1}{q^{-1} + 1 + q} & \frac{1}{\sqrt{q^{-1} + 1 + q}}\\
\frac{1}{\sqrt{q^{-1} + 1 + q}} & \frac{q^{-1} - 1 + q}{q^{-1} + q}\\
\end{pmatrix}
\end{equation}
where $q = e^ {2\pi i/5}$ for the unitary and $q = e^ {4\pi i/5}$ for the non-unitary case which we will call the Fibonacci and Yang-Lee theory respectively. Explicitly, the $F$ symbols are
\begin{equation}
F^{\tau\tau\tau}_{\tau ef} = \frac{1}{\phi}
\begin{pmatrix}
1 & \sqrt{\phi}\\
\sqrt{\phi} & -1\\
\end{pmatrix}
\end{equation}
with $\phi = \frac{1 + \sqrt{5}}{2}$ for the unitary and $\phi' = \frac{1 - \sqrt{5}}{2}$ for the non-unitary case.

\simplesection{Ocneanu tube algebra and central idempotents}
In \cite{Bultinck2017} it was shown that the ground state space of the MPO-injective PEPS is given by the support of the objects $A_{abcd}$ defined by
\begin{equation}
\includegraphics[width=0.5\linewidth]{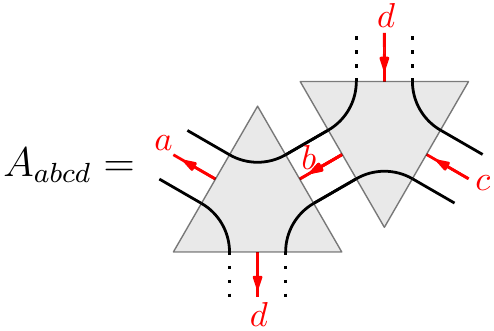}
\end{equation}
when interpreted as matrices from the indices $c$ to $a$. The dotted lines indicate that they are connected, allowing for an arbitrary numbers of MPO tensors along the connecting line. These objects are basis elements of the Ocneanu tube algebra, and for unitary fusion categories, they were shown to form a $C^*$ algebra. This allows the construction of central idempotents which project onto the different topological sectors. When we try to generalise this to non-unitary fusion categories, we find that these objects still form an algebra but are no longer closed under Hermitian conjugation, so that we have no a priori reason to assume the existence of idempotents. Nevertheless, explicit formulas for the idempotents of modular braided categories exist \cite{koenig2010quantum} based on their modular $S$ matrices and the $R$ matrices, solutions to the hexagon equations. The non-unitary $\text{su}(2)_k$ share these properties with their unitary counterparts, and we therefore still expect the existence of idempotents. In this paper we take a pragmatic approach and use a constructive algorithm to find the central idempotents, as described in \cite{Bultinck2017}. For the example of the Yang-Lee theory we find 4 central idempotens, 3 of which are one dimensional and one is 2-dimensional, in agreement with the 7 different allowed tubes. Their explicit forms are
\begin{align}
&\mathcal{P}_{1,1\bar{1}} = -\frac{1}{\phi'\sqrt{5}}\left[A_{1111} + \phi'A_{1\tau1\tau}\right],\nonumber\\[.5em]
&\mathcal{P}_{\tau,\tau\bar{1}} = -\frac{1}{\phi'\sqrt{5}}\left[A_{\tau\tau\tau1} + e^{\frac{2\pi i}{5}}A_{\tau1\tau\tau} + \sqrt{\phi'}e^{\frac{\pi i}{5}}A_{\tau\tau\tau\tau}\right],\nonumber\\[.5em]
&\mathcal{P}_{\tau,1\bar{\tau}} = -\frac{1}{\phi'\sqrt{5}}\left[A_{\tau\tau\tau1} + e^{-\frac{2\pi i}{5}}A_{\tau1\tau\tau} + \sqrt{\phi'}e^{-\frac{\pi i}{5}}A_{\tau\tau\tau\tau}\right],\nonumber\\[.5em]
&\mathcal{P}_{1,\tau \bar{\tau}} = -\frac{1}{\phi'\sqrt{5}}\left[\phi'^2 A_{1111} - \phi'A_{1\tau1\tau}\right],\nonumber\\[.5em]
&\mathcal{P}_{\tau,\tau\bar{\tau}} = -\frac{1}{\phi'\sqrt{5}}\left[\phi'A_{\tau\tau\tau1} + \phi'A_{\tau1\tau\tau} + \frac{1}{\sqrt{\phi'}}A_{\tau\tau\tau\tau}\right].
\end{align}
These idempotents were shown to correspond to anyonic excitations in the MPO injective PEPS. The topological spin associated to these anyons \cite{Bultinck2017}, which is strongly related to the conformal spin, is given by
\begin{equation}
h_{1\bar{1}} = 0, \quad h_{\tau\bar{1}} = -\frac{1}{5}, \quad h_{1\bar{\tau}} = \frac{1}{5}, \quad h_{\tau\bar{\tau}} = 0.
\end{equation}

\simplesection{Strange correlator and Yang-Lee edge singularity}
As mentioned in the introduction, it was shown in \cite{freedman2012galois} that these non-unitary string net wave functions can not have a Hermitian gapped parent Hamiltonian and therefore can not be topological ground states of a Hermitian Hamiltonian.  Our main interest however is in the mapping to partition functions of classical statistical mechanics models, where the non-unitarity translates into models out of equilibrium. We do this by projecting the physical degrees of freedom of the topological PEPS onto a product state, resulting in the partition function of a critical statistical mechanics model descibed by a CFT. We use exactly the same mapping as in \cite{vanhove2018mapping} for the unitary Fibonacci model, and project all the physical degrees of freedom to $\mathbf{\tau}$, resulting in the following Boltzmann weights:
\begin{equation}
\includegraphics[width=0.8\linewidth]{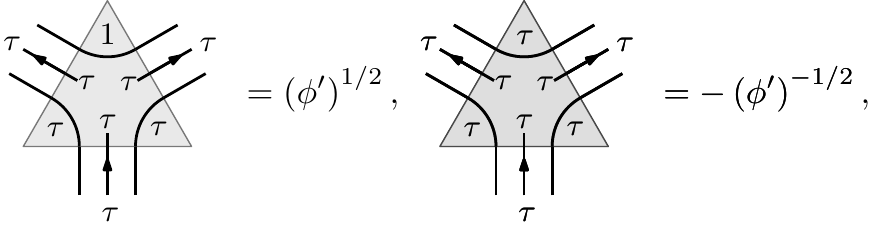}
\end{equation}
including all rotations and reflections of the PEPS tensors. 
\begin{figure}
	\center
	\includegraphics[width=0.475\textwidth]{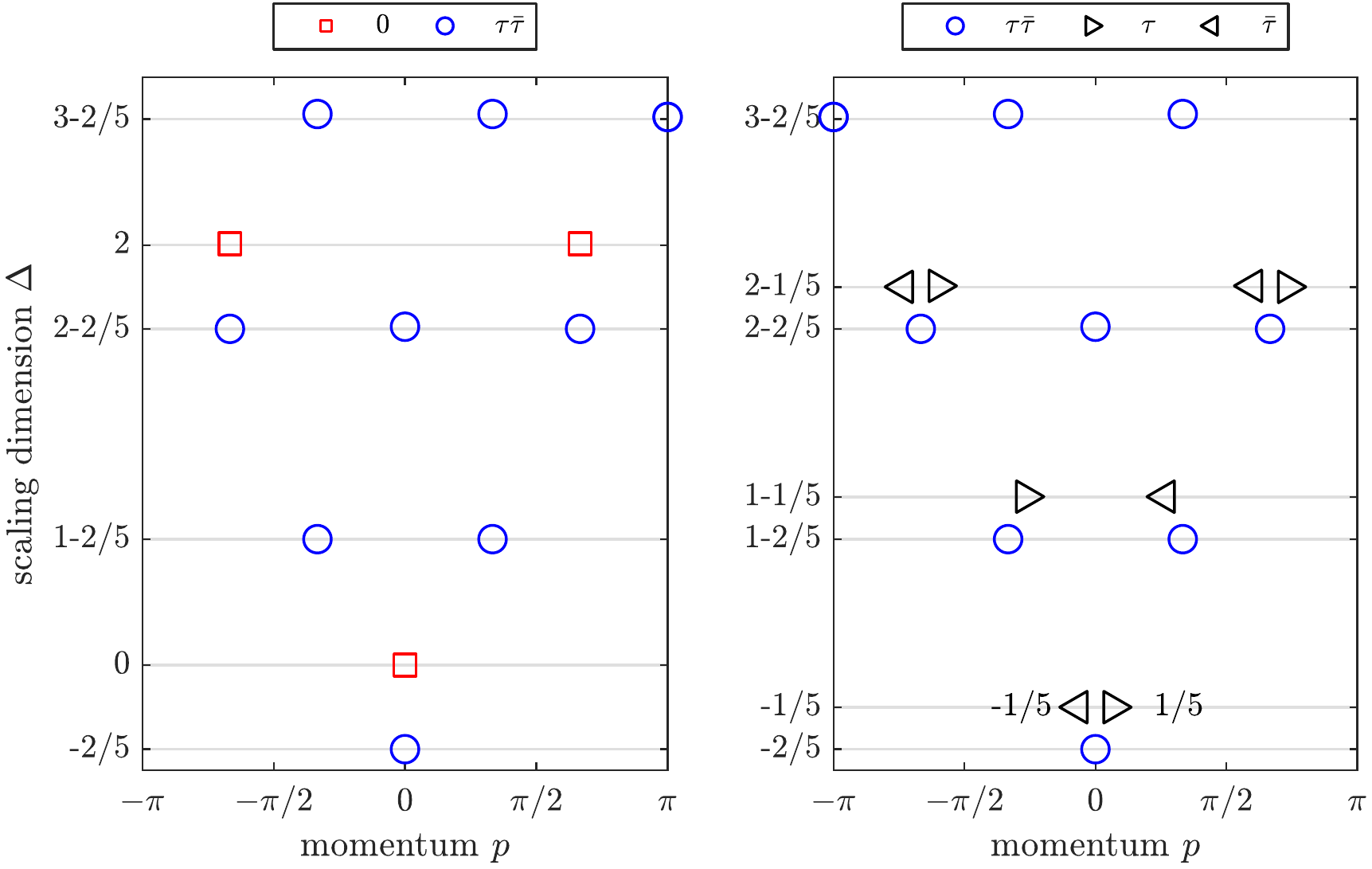}
	\caption{Topological sector labeling of finite-size CFT spectra (scaling dimension $\Delta$ versus momentum) on a cylinder with a circumference of 18 sites with no defect line on the left and a $\tau$ defect line on the right. The exact topological correction to the conformal spin is denoted next to the first appearance of the respective idempotents.}
	\label{spectra}
\end{figure}
This model corresponds to the critical hard-hexagon model with negative Boltzmann weights, with the internal $\{1,\tau\}$-loops corresponding to the precense and absence of a particle in the classical model, respectively. It has long been known \cite{baxter1980hard} that beside a critical fugacity of $z_c = \phi^5$, this model also is critical at $z_c = \left(\phi'\right)^5$, which is precisely the value we recover from the non-unitary PEPS tensors after applying the strange correlator. The resulting partition function on a cylinder can be diagonalised with or without a flux insertion ($\tau$ defect), and the spectra can be projected onto the topological sectors using the idempotents discussed above. The results for the trival flux and $\tau$ flux are shown in Figure \ref{spectra}, where the positivity of the eigenvalues despite a non-Hermitian transfer matrix allows for the identification with the conformal spectrum of the Yang-Lee edge singularity. We can make the following correspondence between the topological sectors and the Virasoro characters of the Yang-Lee edge singularity:
\begin{align}
	\mathcal{P}_{1,1\bar{1}} &\rightarrow \chi_0\chi_0^*,\nonumber\\
	\mathcal{P}_{\tau,\tau\bar{1}} &\rightarrow \chi_{-\frac{1}{5}}\chi_0^*,\nonumber\\
	\mathcal{P}_{\tau,1\bar{\tau}} &\rightarrow \chi_0\chi_{-\frac{1}{5}}^*,\nonumber\\
	\mathcal{P}_{1,\tau\bar{\tau}} + \mathcal{P}_{\tau,\tau\bar{\tau}} &\rightarrow \chi_{-\frac{1}{5}}\chi_{-\frac{1}{5}}^*.
\end{align}
We note that the conformal spectrum without defect has already been obtained in \cite{ardonne2011microscopic} in the context of non-unitary anyonic spin chains, where it was found that the Hamiltonian for these systems possesses a topological symmetry. In the MPO formalism, the local terms of the Hamiltonian of these anyonic spin chains can be written in terms of PEPS tensors as
\begin{equation}
\includegraphics[width=0.3\linewidth]{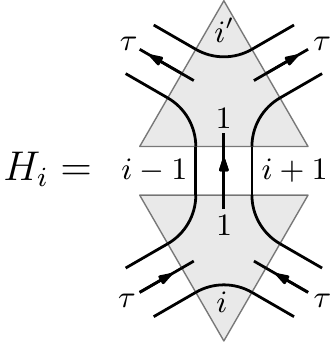}
\end{equation}
where the topological MPO symmetry is manifest because of the pulling-through equation.

\simplesection{Conclusions}
We have generalised the matrix product operator formalism for string-net wavefunctions to non-unitary solutions of the pentagon equation, provided the underlying category is a braided modular tensor fusion category. We have argued that the construction of the topological superselection sectors (anyons) can still be carried out even though the tube algebra is no longer closed under Hermitian conjugation. This was illustrated numerically by applying the strange correlator map to the Yang-Lee model and obtaining conformal spectra with the explicit identification of the sectors. It would be interesting to further generalise the strange correlator to logarithmic CFT. The lattice description of the TQFT-CFT correspondence for these CFTs promises to shed new light on old problems such as percolation, which is believed to be described by a logarithmic CFT in the continuum limit.

\begin{acknowledgments}
We would like to thank Dominic Williamson and Nick Bultinck for insightful conversations. This work is supported by an Odysseus grant from the FWO, ERC grants QUTE (647905) and ERQUAF (715861), the EU grant SIQS and FWO PhD-grant (R.V).
\end{acknowledgments}

\bibliographystyle{apsrev4-1}
\bibliography{non_unitary.bib}

\end{document}